\documentclass{aa}  

\usepackage{graphicx}
\usepackage{txfonts}
\usepackage{lipsum}
\usepackage{subcaption}         % necessary for continued figures, example in section 3
\usepackage{lscape}             % to rotate a single page table, example in appendix.
\usepackage{placeins}           % useful with \FloatBarrier, to keep 
\makeatletter
\let\linenumbers\relax
\makeatother

\begin{document}
%%%%%%%%%%%%%%%%%%%%%%%%%%%%%%%%%%%%%%%%
% if you use custom commands in your title,
% ensure to check your title when submitting!
%%%%%%%%%%%%%%%%%%%%%%%%%%%%%%%%%%%%%%%%
   \title{Mitigating variance shrinkage in multi-task galaxy property inference from imaging data via statistical manifold alignment}

   %\subtitle{Subtitle}
   \titlerunning{Mitigating variance shrinkage via statistical manifold alignment}
%%%%%%%%%%%%%%%%%%%%%%%%%%%%%%%%%%%%%%%%
% Please separate each author with the \and command
%
% Use the \corrauth to provide the corresponding
% author address. It will be automatically inserted as 
% footnote in the PDF output.
%
% Please DO NOT include ORCIDs next to author names.
% Instead, please provide an active address for each coauthor:
% it will be automatically extracted by EDPS editorial system, 
% and co-authors will be be able to authenticate their ORCID.
%
% Only authenticated ORCIDs will be taken into account.
% ORCIDs included here will be removed.
%%%%%%%%%%%%%%%%%%%%%%%%%%%%%%%%%%%%%%%%

   \author{Zhijian Luo\inst{1}\corrauth{zjluo@shnu.edu.cn}       
        \and Jianzhen Chen\inst{1}\corrauth{jzchen@shnu.edu.cn} 
        }

   \institute{Shanghai Key Lab for Astrophysics, Shanghai Normal University, 200234, Shanghai, People’s Republic of China}
   %\date{Received September 30, 20XX}

% \abstract{}{}{}{}{}
% 5 {} token are mandatory
 
  \abstract
   {Inferring photometric redshifts ($z$) and key stellar population properties ($\log(M_*/\mathrm{M}_\odot)$, $D_n4000$, and $\mathrm{H}\delta_A$) directly from multi-band imaging is increasingly critical for spectroscopically limited large surveys such as LSST and CSST. However, conventional point-wise loss functions (e.g., mean squared error or Huber loss) employed in deep neural networks often induce ``variance shrinkage''---a regression-to-the-mean effect that contracts predicted variances, distorts non-linear physical manifolds (such as bimodal distributions), and truncates heavy-tail populations. Here, we present ViTMultiTask, a vision transformer framework that incorporates a multi-scale maximum mean discrepancy (MMD) loss to enforce higher-order joint distribution alignment in reproducing kernel Hilbert space (RKHS). Trained on 83,846 galaxies from Subaru HSC five-band ($g,r,i,z,y$) imaging cross-matched with DESI Data Release 1 Bright Galaxy Survey spectra, the model achieves high point-estimate precision on an independent test set of 5,000 galaxies: $\sigma_{\mathrm{NMAD}} = 0.020$ for $z$, $0.169\,\mathrm{dex}$ for $\log(M_*/\mathrm{M}_\odot)$, $0.073$ for $D_n4000$, and $1.185\,\mathrm{\AA}$ for $\mathrm{H}\delta_A$. Ablation experiments demonstrate that incorporating the MMD loss effectively mitigates variance shrinkage and more accurately reproduces the observed spectroscopic parameter distributions (including bimodal structures and heavy tails), while maintaining point-estimate accuracy statistically consistent with the baseline (RMSE changes $\le 3.5\%$) and catastrophic outlier rates fully comparable to the baseline. This kernel-based manifold alignment offers a statistically grounded and practical paradigm for comprehensive galaxy property inference in the deep imaging era.
   }
   

   \maketitle

%%%%%%%%%%%%%%%%%%%%%%%%%%%%%%%%%%%%%%%%%%%%%%%%%%%%%%%%%%%%%%
\section{Introduction} \label{sec:intro}

Accurately determining the physical properties of galaxies for large statistical samples is fundamental to understanding cosmic structure formation and galaxy evolution \citep{2013ARA&A..51..393C,2014ARA&A..52..415M}. Next-generation wide-field imaging surveys---including the Vera C. Rubin Observatory's Legacy Survey of Space and Time (LSST; \citealt{2019ApJ...873..111I}), the Euclid space telescope \citep{2011arXiv1110.3193L,2025A&A...697A...1E}, and the Chinese Space Station Telescope (CSST; \citealt{zhan2011consideration,2018MNRAS.480.2178C,2026SCPMA..6939501C})---will acquire imaging data for billions of galaxies. This vastly exceeds the spectroscopic follow-up capacity of contemporary spectroscopic surveys such as the Dark Energy Spectroscopic Instrument (DESI; \citealt{2022AJ....164..207D,2024AJ....167...62D,2024AJ....168...58D}). Consequently, directly inferring photometric redshift ($z$) and key stellar population evolutionary parameters---total stellar mass ($\log(M_*/\mathrm{M}_\odot)$), $D_n4000$ break index, and $\mathrm{H}\delta_A$ equivalent width---from pixel-level multiband images has become an important objective bridging large-scale photometric surveys and galaxy evolution theory. 

Traditional methods fall broadly into two categories: spectral energy distribution (SED) template fitting (e.g., LePhare, EAZY; \citealt{1999MNRAS.310..540A,2008ApJ...686.1503B}) and classical machine learning algorithms (e.g., random forests, self-organizing maps; \citealt{2021MNRAS.502.2770M,2024MNRAS.52712140L,2024ApJ...967...60J}). Template fitting is limited by template library incompleteness and parameter degeneracies in non-typical galaxies \citep{2020ApJ...904...33L,2023ApJ...944..141P}; classical machine learning methods rely on tabular inputs such as integrated photometric magnitudes or colors, discarding rich spatial morphological and dynamical information---such as concentration, bulge-to-disk ratio, and tidal structures---that are physically correlated with galaxy evolutionary stages \citep{2003MNRAS.341...33K,2014ARA&A..52..291C}. Deep learning methods represented by convolutional neural networks (CNNs) have achieved end-to-end inference directly from image cutouts \citep{2018A&A...609A.111D,2019A&A...621A..26P}. However, CNNs are restricted by local receptive fields, posing inherent limitations in capturing long-range cross-band dependencies and cross-scale spatial correlations. Vision transformers (ViT; \citealt{2020arXiv201011929D,2021arXiv211001024Y,2026ApJS..283...62L}), leveraging global self-attention mechanisms, can efficiently compute global relationships between image patches, providing a superior architectural choice to simultaneously probe subtle spatial structures and multi-color spatial gradients in galaxies.

Although deep neural networks have demonstrated good precision in mapping multiband pixel fluxes to scalar galaxy properties, most existing approaches still treat individual physical parameters as independent single-task regression problems and optimize them using point-wise loss functions, such as mean squared error (MSE) or Huber loss. This paradigm has two potential limitations. First, processing tasks independently makes it difficult to fully share underlying representations across parameters, which may lead to internally inconsistent predictions and thus affect the recovery of physical scaling relations, such as the star-forming main sequence \citep{2014ApJS..214...15S} or the mass--metallicity relation \citep{2004ApJ...613..898T}. Second, and more critically, point-wise loss functions tend to induce variance shrinkage (i.e., regression to the mean): to minimize mean squared residuals under observational noise, the network tends to predict the conditional expectation $\mathbb{E}[Y \mid X]$. By Jensen's inequality, the variance of the conditional expectation is upper-bounded by the total variance, $\mathrm{Var}(\mathbb{E}[Y \mid X]) \le \mathrm{Var}(Y)$. This can artificially narrow the predicted distribution, potentially suppressing heavy tails (e.g., low-mass or high-redshift galaxies) and distorting astrophysical bimodal structures, such as the transition between the star-forming blue cloud and the quiescent red sequence \citep{2001AJ....122.1861S, 2004ApJ...600..681B}. For statistical cosmology applications---constructing the galaxy stellar mass function (GSMF; \citealt{2013ApJ...777...18M}), tracing star-formation quenching histories \citep{2010ApJ...721..193P}, or selecting rare extreme objects---preserving overall distribution fidelity and physical boundaries is often more important than minimizing residuals for individual objects. Generative models, such as normalizing flows \citep{2015arXiv150505770J,2026ApJ...997...65R}, can model full probability density functions, but they generally incur substantial computational overhead and training complexity when scaled to high-dimensional multi-task parameter spaces.

To overcome these limitations without the heavy computational burden of generative models, non-parametric statistical distances provide a useful alternative for encouraging distribution alignment. Distribution alignment methods, especially those developed in the domain adaptation (DA) literature, have gained increasing traction in astrophysics for mitigating dataset shifts and cross-survey domain gaps \citep{2021MNRAS.506..677C,2022MLS&T...3c5007C,2023MLS&T...4b5013C}. Such methods have been applied to tasks including galaxy merger classification \citep{2021MNRAS.506..677C}, semi-supervised universal domain adaptation for cross-survey morphology classification and anomaly detection (DeepAstroUDA; \citealt{2023MLS&T...4b5013C}), and strong gravitational lensing modeling \citep{2024arXiv241103334A}. Whereas much of this work focuses on inter-domain transfer---aligning feature representations across distinct datasets---the same distribution-alignment principle can be adapted to improve distributional fidelity within a single survey sample. In our setting, aligning the predicted joint parameter distribution with the distribution of observed spectroscopic labels can help mitigate the variance shrinkage and tail underrepresentation associated with pointwise regression losses.

Among the statistical distances and divergences used in the DA literature---such as optimal transport  (OT; \citealt{peyré2020computationaloptimaltransport}), especially the Wasserstein distance, and Kullback--Leibler (KL) divergence---maximum mean discrepancy (MMD; \citealt{gretton2012optimal,2021arXiv211015073S}) offers practical advantages for multivariate parameter regression. Unlike KL objectives that rely on explicit density estimation, MMD avoids explicit mini-batch density estimation, which is often unstable for small batches or in higher dimensions. Compared with OT, which typically requires solving a linear program or Sinkhorn iterations at every training step, MMD has a closed-form kernel-based estimator that can be integrated directly into standard GPU mini-batch gradient descent, and it incurs lower computational overhead. Moreover, a mixture of Gaussian kernels at multiple scales allows MMD to penalize both local manifold discrepancies and broader distributional mismatches. This makes MMD a computationally efficient, stable, and smooth regularization term for better reproducing the breadth and bimodal or multimodal structure of the observed target distribution during continuous multi-task training.

In this paper, we propose ViTMultiTask---a statistically grounded multi-task learning framework that jointly infers $z$, $\log(M_*/\mathrm{M}_\odot)$, $D_n4000$, and $\mathrm{H}\delta_A$ directly from multiband photometric images. By introducing task-specific tokens into a vision transformer backbone and incorporating a multi-scale MMD penalty, our framework encourages the predicted joint distribution to align with the empirical joint distribution of the spectroscopic labels in a reproducing kernel Hilbert space (RKHS; \citealt{2010arXiv1003.0887S}). We train ViTMultiTask on 83,846 five-band ($g, r, i, z, y$) Subaru Hyper Suprime-Cam (HSC) image cutouts cross-matched with high-quality spectroscopic labels from the Dark Energy Spectroscopic Instrument (DESI; \citealt{2023AJ....165..253H}) Data Release 1 Bright Galaxy Survey (BGS), and show that MMD regularization effectively mitigates variance shrinkage, more closely reproducing the observed distributions of the spectroscopic labels (including bimodal structures and heavy-tailed populations) and maintaining catastrophic outlier rates comparable to those of the baseline, with only a marginal cost in point-estimate root mean square error (RMSE).

This paper is structured as follows: Section 2 describes the observational data and preprocessing pipeline; Section 3 details the ViTMultiTask architecture and the MMD-constrained loss function; Section 4 presents inference performance, distribution recovery, and ablation study results; Section 5 summarizes the main conclusions and outlines future applications to CSST, Euclid, and LSST.

\section{Data and sample construction}
\label{sec:data}

To assemble a benchmark dataset containing both multiband astronomical imaging and high-fidelity physical property labels, we cross-match and curate samples from two recent data products derived from the DESI Data Release 1 \citep{2026AJ....171..285D}. The imaging components are sourced from the GalaxiesML-Spectra dataset\footnote{\url{https://zenodo.org/records/16989593}} \citep{2024arXiv241000271D,2025arXiv251022527H}, which spatially cross-matches spectroscopic targets from DESI Data Release 1 with five-band ($g, r, i, z, y$) imaging from the public data release 2 (PDR2; \citealt{2019PASJ...71..114A}) of Hyper Suprime-Cam Subaru Strategic Program (HSC-SSP; \citealt{2018PASJ...70S...4A}). HSC-SSP offers exceptionally deep optical imaging with superb spatial resolution (median seeing $\sim 0\farcs6$) and high signal-to-noise ratio across all bands, enabling the extraction of fine morphological structures and color gradients essential for stellar population inferences. The initial base catalog was constructed by selecting targets with clean spectroscopic redshifts ($\text{ZWARN} = 0$; \citealt{2023AJ....165..253H}), filtering out inconsistent redshifts and duplicate observations, aligning within a $0\farcs5$ matching radius, and excluding stars classified by DESI, yielding a foundational sample of 134,533 galaxy images. Although the original GalaxiesML-Spectra release includes one-dimensional spectra and metadata, we strictly utilize its $64 \times 64$ pixel five-band HSC image cutouts and unique target identifiers (\text{TARGETID}) for catalog cross-matching.

Ground-truth astrophysical labels are derived from the DESI Data Release 1 BGS \citep{2023AJ....165..253H} Value-Added Catalog \footnote{\url{http://www.doi.org/10.57760/sciencedb.35360}} (VAC; \citealt{2026arXiv260719162L}) , which covers approximately 6.7 million galaxy spectra. This VAC executes systematic stellar population modeling via a full-spectral fitting pipeline, deriving comprehensive physical parameters including spectroscopic redshift ($z$), total stellar mass ($M_*$), light-weighted stellar age ($\mathrm{age}_L$), stellar metallicity ($Z_L$), stellar velocity dispersion ($\sigma_*$), dust attenuation $E(B-V)$, spectral indices (e.g., $D_n 4000$ break and $\mathrm{H}\delta_A$ Balmer absorption line equivalent width), as well as flux and equivalent widths of major optical emission lines.

We cross-match the HSC imaging dataset with the DESI Data Release 1 BGS VAC using the unique \texttt{TARGETID}, securing 84,284 initial matches. To ensure physical reliability and eliminate non-physical outliers, we apply range constraints on the derived parameters: $z \le 0.8$, $10^7 \le M_*/\mathrm{M}_\odot \le 10^{13}$ (i.e., $\log(M_*/\mathrm{M}_\odot) \in [7, 13]$), $0.8 \le D_n 4000 \le 2.5$, and $-5.0 \le \mathrm{H}\delta_A \le 10.0~\mathrm{\AA}$. This quality cut discards only 438 unphysical anomalies ($\sim 0.5\%$ rejection rate), resulting in a clean dataset of 83,846 galaxies. Each object contains high-resolution $g, r, i, z, y$ imaging alongside robust spectroscopic labels ($z$, $\log(M_*/\mathrm{M}_\odot)$, $D_n 4000$, and $\mathrm{H}\delta_A$).

We prioritize $z$, $\log(M_*/\mathrm{M}_\odot)$, $D_n 4000$, and $\mathrm{H}\delta_A$ as our primary target variables, which together form a multi-dimensional diagnostic system characterizing galaxy state and evolutionary pathway \citep{2003MNRAS.341...33K, 2007PASJ...59..541L, 2025arXiv251025216L}. $z$ establishes the cosmological distance scale and rest-frame spectral coverage; $M_*$ governs the depth of the gravitational potential well and long-term evolutionary trajectory; $D_n 4000$ is sensitive to old stellar populations and traces mean light-weighted stellar age, though it is also degenerate with metallicity; and $\mathrm{H}\delta_A$ equivalent width spikes violently within $0.1\text{--}1~\mathrm{Gyr}$ following a starburst, acting as a crucial tracer of recent star formation activity and rapid quenching episodes.

Direct inference of these high-dimensional parameters from multiband imaging is supported by a well-established physical basis. Broadband colors (e.g., $g-r$, $r-z$) offer constraints on low-resolution SEDs, which are generally informative for estimating $z$, $M_*$, and $D_n 4000$. At the same time, image cutouts resolve fine spatial features---such as angular size, concentration, radial color gradients, and structural perturbations \citep{2014ARA&A..52..291C}---that are often physically linked to star formation history and violent evolutionary events (e.g., core structural compaction frequently accompanies enhanced $\mathrm{H}\delta_A$ absorption; \citealt{2014MNRAS.438.1870D}). Deep neural networks are capable of mapping these nonlinear pixel-level spatial-color patterns to small-scale spectroscopic diagnostics \citep{2021arXiv211101154B,2023MNRAS.520.3529E}, thereby facilitating cross-modal parameter inference.

\begin{figure*}[t]
\centering
\includegraphics[width=\textwidth]{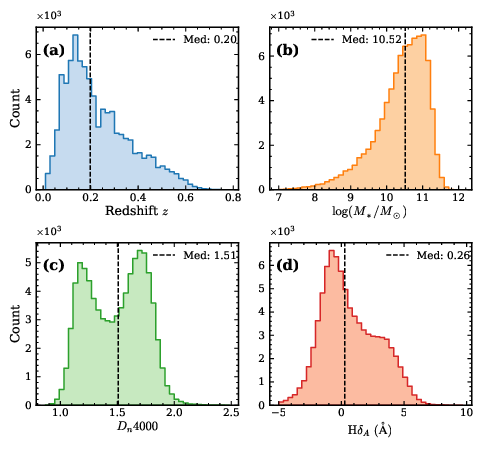}
\caption{ Parameter distributions of the four target physical properties across the galaxy sample: (a) redshift $z$; (b) stellar mass $\log(M_*/\mathrm{M}_\odot)$; (c) $D_n 4000$ index; and (d) $\mathrm{H}\delta_A$ absorption line equivalent width. Vertical dashed black lines in each panel indicate the median value of the corresponding parameter.}
\label{fig:para_dist}
\end{figure*}

As illustrated in Figure Fig.~\ref{fig:para_dist}, the final 83,846 galaxies span a broad and representative parameter space: redshifts are concentrated within $z \in [0.05, 0.60]$, consistent with the selection function of the low-redshift DESI BGS; stellar mass follows a slightly left-skewed, near-Gaussian distribution with a median of $\log(M_*/\mathrm{M}_\odot) = 10.52$. Furthermore, $D_n 4000$ exhibits a prominent bimodal distribution, whereas $\mathrm{H}\delta_A$ displays a right-skewed heavy tail, jointly reflecting a diverse population comprising star-forming blue galaxies, green-valley transition systems, and quenched red sequence galaxies. For neural network training and evaluation, we randomly partition the dataset into training (73,846 galaxies), validation (5,000 galaxies), and test (5,000 galaxies) subsets, corresponding to an approximate ratio of $88:6:6$. This partitioning strategy maximizes the training sample size required to effectively optimize the high-capacity vision transformer backbone, while reserving a test set large enough for overall performance estimates. By adopting uniform random sampling, the test set is, by construction, an unbiased random subset of the parent population. A comparison of the parent and test distributions shows no significant bias beyond expected sampling fluctuations across the main parameter ranges and the targeted tail regions, although the absolute counts in the extreme tails remain limited by the parent sample size.

\section{Methodology and training pipeline}
\label{sec:method}

\subsection{Data Preprocessing and Label Standardization}
\label{sec:preprocessing}

Before model training, we applied a rigorous data preprocessing and standardization pipeline to the multiband Subaru HSC images and their corresponding spectroscopic physical labels. Notably, no additional per-channel scaling or cross-band normalization was applied to the images, so as to directly preserve the intrinsic physical flux ratios and color gradient information across bands. This choice was made to retain the inherent color diagnostic power, while the input dynamic range is handled by the network's adaptive weight mechanisms during training.

For the physical labels, given that stellar mass spans several orders of magnitude, we first applied a logarithmic transformation, $\log_{10}(M_*/\mathrm{M}_\odot)$, to compress its dynamic range. In contrast, redshift $z$ (typically $0 < z \lesssim 0.8$ in our sample), the $D_n 4000$ index, and the $\mathrm{H}\delta_A$ equivalent width are intrinsically linear-scale quantities with relatively compact empirical distributions, thus requiring no additional non-linear compression. 

To strictly prevent data leakage, we first split the full set of 83,846 samples into training, validation, and test sets, containing 73,846, 5,000, and 5,000 samples, respectively (approximately 88\%, 6.0\%, and 6.0\% of the total). The validation set was used solely for hyperparameter tuning and selection of the best model weights, while the test set was reserved exclusively for final inference evaluation; neither participated in gradient updates during training. Subsequently, we computed the mean and standard deviation of the four physical labels exclusively from the training set samples, and applied $Z$-score standardization:
\begin{equation}
y_{\mathrm{norm}, k} = \frac{y_k - \mu_k}{\sigma_k}, \quad k \in \{z, \log_{10}(M_*/\mathrm{M}_\odot), D_n 4000, \mathrm{H}\delta_A\},
\end{equation}
where $\mu_k$ and $\sigma_k$ are the mean and standard deviation of the $k$-th label in the training set. This standardization removes the interference caused by different numerical scales among the physical targets on multi-task gradient balancing, and significantly improves the convergence stability of multi-task learning.

For the validation and test sets, we strictly applied the same training-set statistics for normalization and inverse transformation ($\hat{y}_k = \hat{y}_{\mathrm{norm}, k} \cdot \sigma_k + \mu_k$) during inference, ensuring strict self-consistency across cross-dataset evaluation.

\subsection{ViTMultiTask architecture and task tokens}
\label{sec:architecture}

\begin{figure*}[t]
\centering
\includegraphics[width=\textwidth]{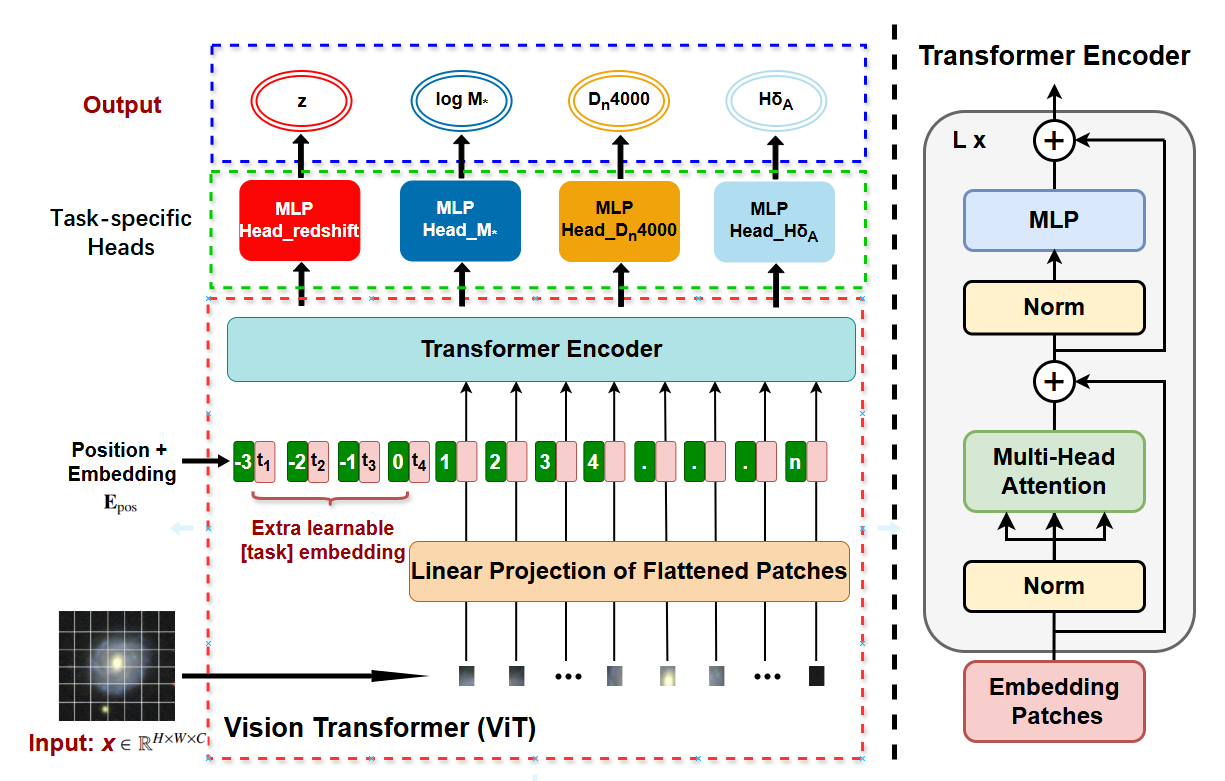}
\caption{Architecture of the multi-task galaxy parameter prediction network (ViTMultiTask). \textbf{Left:} Overall multi-task prediction workflow. After mapping the multiband image cutout $\mathbf{x} \in \mathbb{R}^{H \times W \times C}$ into patch embeddings, four learnable task tokens ($t_1\text{--}t_4$) are prepended to the sequence and element-wise added with positional embeddings $\mathbf{E}_{\mathrm{pos}}$ before being fed into the transformer encoder backbone for feature extraction and representation decoupling. Finally, the four decoupled task tokens are routed into their respective heterogeneous MLP prediction heads to concurrently infer $z$, $\log(M_*/\mathrm{M}_\odot)$, $D_n 4000$, and $\mathrm{H}\delta_A$. 
\textbf{Right:} Internal structure of a single transformer encoder block, comprising layer normalization, multi-head attention, and MLP feed-forward layers with residual connections, stacked across $L = 8$ layers.}
\label{fig:arch}
\end{figure*}

We construct ViTMultiTask, a deep multi-task network integrating a vision transformer (ViT; \citealt{2020arXiv201011929D}) backbone with a task-specific token mechanism (Fig.~\ref{fig:arch}). The overall architecture proceeds from input embedding through transformer encoding to task-specific prediction.

Taking $64 \times 64 \times 5$ image cutouts as input, the backbone first maps image patches into sequence embeddings via linear projection of flattened patches. In practice, this is implemented using $4 \times 4$ convolutional kernels with a stride of 4 and zero-padding, which downsamples the spatial dimensions to $16 \times 16$ and yields 256 spatial patch vectors with an embedding dimension of $d = 80$.

To effectively decouple features across the four target variables and eliminate the need for global average pooling, we introduce four learnable task-specific tokens $\mathbf{T}_{\mathrm{task}} \in \mathbb{R}^{4 \times 80}$, a strategy inspired by recent advances in multi-task vision transformers \citep{2024ITNNL..35.9579T}. These tokens are prepended to the spatial patch sequence, forming an input sequence of length $256 + 4 = 260$. A learnable 1D position embedding $\mathbf{E}_{\mathrm{pos}} \in \mathbb{R}^{260 \times 80}$ is added across the entire sequence, enabling the task tokens to attend to spatial semantics from the first layer onward.

The backbone comprises $L = 8$ transformer blocks. Each block incorporates a 4-head self-attention module with a head dimension of $d_{\mathrm{head}} = 20$, an MLP feed-forward network with an expansion factor of 4 (hidden dimension 320), layer normalization, and dropout ($p = 0.2$). Through multi-layer self-attention interactions, the four task tokens at the sequence head are dynamically updated by aggregating information from all spatial patches, effectively decoupling representations for each target without spatial compression.

After the final transformer layer, these task tokens are extracted and routed to task-specific prediction heads. The heads adopt heterogeneous architectures tailored to the distinct physical characteristics of each target variable. For global, scale-dependent quantities—redshift and stellar mass—which are primarily governed by overall SED shapes and galaxy sizes, we employ two-layer fully connected networks with dimensions 256 and 192, using ReLU activations \citep{glorot2011deep} and layer normalization \citep{2016arXiv160706450L}.

For $D_n4000$, which probes the $4000~\mathrm{\AA}$ break and is sensitive to stellar population age, we use a $[256, 128]$ architecture with GELU activations \citep{2016arXiv160608415H} and batch normalization \citep{2015arXiv150203167I}, leveraging mini-batch statistics to enhance sensitivity to break strength. Finally, to extract subtle absorption-line signatures from broadband imaging, the $\mathrm{H}\delta_A$ head adopts a deeper three-layer structure ($512, 256, 128$) with GELU activations and dual batch normalization, providing the high non-linear capacity needed to resolve signatures of recent starbursts and rapid quenching episodes.

\subsection{Joint loss function with MMD manifold alignment}
\label{sec:loss}

In galaxy astrophysics, the four target parameters are not independent scalar variables; rather, they are intrinsically coupled through stellar population synthesis (SPS; \citealt{2003MNRAS.344.1000B}) and star formation histories (SFHs). Specifically, $D_n4000$ and $\mathrm{H}\delta_A$ trace a tight, non-linear evolutionary manifold \citep{2003MNRAS.341...33K}: $\mathrm{H}\delta_A$ rapidly peaks in post-starburst galaxies before decaying as the population ages, while $D_n4000$ generally rises with mean stellar age (though it remains degenerate with metallicity). Concurrently, $\log(M_*/\mathrm{M}_\odot)$ dictates the gravitational potential depth and gas retention, while $z$ sets the cosmic epoch and redshifted spectral window, jointly modulating the evolutionary timescale and flux normalization of the SED. Jointly modeling these parameters ensures that ViTMultiTask respects this intrinsic astrophysical manifold, preventing the non-physical decoupling that typically arises from independent single-target regressions.

Standard point-wise loss functions (e.g., MSE or Huber loss alone) treat target variables independently and ignore cross-variable covariances, inducing variance shrinkage and over-smoothing of rare evolutionary phases. To improve fidelity to the observed spectroscopic-label distribution, we construct a joint loss function combining a weighted point-wise Huber loss with a maximum mean discrepancy (MMD; \citealt{gretton2012optimal}) distribution alignment penalty. All losses are computed in the standardized target space (see Section~3.1) to ensure that the MMD kernel operates on unit-variance dimensions:
\begin{equation}
\mathcal{L}_{\mathrm{total}} = \sum_{k=1}^{4} w_k \cdot \mathcal{L}_{\mathrm{Huber}}(y_{\mathrm{true}, k}, \hat{y}_{k}) + \lambda_{\mathrm{MMD}} \cdot \mathcal{L}_{\mathrm{MMD}},
\end{equation}
where $y_{\mathrm{true}, k}$ and $\hat{y}_{k}$ denote the standardized spectroscopic labels and predicted values ($y_{\mathrm{norm}, k}$), respectively, and $\mathcal{L}_{\mathrm{Huber}}$ (with threshold $\delta = 1.0$) ensures point-estimate robustness against labeling noise and outliers. The task weight vector is set to $\boldsymbol{w} = [1.0, 1.0, 2.0, 3.0]$ for $z$, $\log(M_*/\mathrm{M}_\odot)$, $D_n4000$, and $\mathrm{H}\delta_A$, respectively. The higher weights assigned to $D_n4000$ and $\mathrm{H}\delta_A$ reflect their relatively smaller dynamic ranges and lower per-instance signal-to-noise ratios compared to redshift and mass, ensuring that the gradient contribution from these spectroscopically sensitive indices is not dominated by the more readily predicted global parameters.

The MMD term penalizes discrepancies between the predicted and observed joint distributions in a reproducing kernel Hilbert space (RKHS; \citealt{2010arXiv1003.0887S}), acting on the standardized four-dimensional target space $\mathbf{y} = [y_1, y_2, y_3, y_4]^\top \in \mathbb{R}^4$. Here, the MMD term is applied to the empirical distribution of the spectroscopic labels, \(P_{\mathrm{obs}}(\mathbf{y})\), rather than to an unobservable intrinsic physical distribution \(P_{\mathrm{int}}(\mathbf{y})\). Because these labels carry measurement uncertainties, spectral-fitting systematics, and selection effects, agreement with \(P_{\mathrm{obs}}(\mathbf{y})\) should be interpreted as improved fidelity to the observed spectroscopic-label distribution, not as exact recovery of the underlying physical distribution. Specifically, we minimize the squared MMD:
\begin{equation}
\mathcal{L}_{\mathrm{MMD}} = \mathbb{E}[K(\hat{\mathbf{y}}, \hat{\mathbf{y}}')] + \mathbb{E}[K(\mathbf{y}, \mathbf{y}')] - 2\mathbb{E}[K(\hat{\mathbf{y}}, \mathbf{y})],
\end{equation}
where $\mathbf{y}'$ and $\hat{\mathbf{y}}'$ denote independent and identically distributed (i.i.d.) draws from the respective target distributions. We adopt a multi-scale Gaussian radial basis function (RBF) mixture kernel $K(\mathbf{u}, \mathbf{v}) = \sum_{\sigma \in \Sigma} \exp\left(-\|\mathbf{u} - \mathbf{v}\|^2 / 2\sigma^2\right)$ with bandwidths $\Sigma = \{0.1, 0.5, 1.0, 2.0, 5.0\}$. Because all physical targets are standardized to unit variance prior to loss computation, the values in $\Sigma$ are expressed in units of the standard deviation of each target dimension. The chosen range ($0.1$ to $5.0$) spans scales from fine-grained feature structures to broad population-level variations, providing comprehensive multi-scale coverage and reducing the need for dataset-specific kernel tuning. By constraining higher-order statistical moments, MMD preserves physical covariances—such as the non-linear $D_n4000$–$\mathrm{H}\delta_A$ evolutionary sequence—that would otherwise be smoothed out by point-wise losses alone.

To determine the MMD loss weight, we performed a sensitivity analysis over $\lambda_{\mathrm{MMD}} \in \{0.1, 0.5, 1.0, 2.0, 5.0\}$. The results indicate that $\lambda_{\mathrm{MMD}} \le 0.5$ is insufficient to mitigate variance shrinkage, whereas $\lambda_{\mathrm{MMD}} \ge 2.0$ may over-emphasize global distribution alignment at the cost of increased point-wise RMSE penalty ($>5\%$). We therefore set $\lambda_{\mathrm{MMD}} = 1.0$, which provides a favorable balance between point-estimate precision and distribution alignment.

\subsection{Training Protocol and Data Augmentation}
\label{sec:training}

The network is implemented in TensorFlow and trained on a single NVIDIA L40S GPU. To improve generalization and enforce robustness against the arbitrary orientation of galaxies on the sky, we apply spatial symmetry augmentations during training, including random horizontal and vertical flips alongside $90^\circ$ rotations. These operations correspond to the $D_4$ dihedral symmetry group and encourage the model to learn orientation-invariant representations, preventing overfitting to particular morphological alignments in the sky plane.

We train the model using the Adam optimizer \citep{2014arXiv1412.6980K} with a batch size of 256 for a total of 300 epochs, which completes in approximately 2.5 hours on our hardware setup. With a training set size of 73,846 samples, each epoch comprises approximately 289 iterations. The initial learning rate is set to $1 \times 10^{-4}$ and decays following a cosine schedule over the first 50,000 steps (corresponding to roughly 173 epochs) down to a terminal value of $1 \times 10^{-6}$ (1\% of the initial rate). For the remaining 127 epochs, the learning rate is held constant at $1 \times 10^{-6}$ to refine convergence near the loss minimum.

The validation loss is monitored after each epoch, and the model checkpoint achieving the lowest validation loss is retained for final evaluation. During inference, the normalized predictions are transformed back to physical units using the training-set mean and standard deviation $\{\mu_k, \sigma_k\}_{k=1}^4$, strictly maintaining consistency with the label standardization protocol described in Section~\ref{sec:preprocessing}.

\section{Results and Astrophysical Discussion}
\label{sec:results}

\subsection{Evaluation Metrics}
\label{sec:metrics}

To evaluate the predictive accuracy and distributional fidelity of our multi-task pipeline, we assess the model on an independent test set of $N = 5{,}000$ galaxies. For physical parameters $k \in \{\log(M_*/\mathrm{M}_\odot), D_n 4000, \mathrm{H}\delta_A\}$, we define the prediction residual for the $i$-th galaxy as $\Delta y_{i,k} = \hat{y}_{i,k} - y_{i,k}$, where $y_{i,k}$ and $\hat{y}_{i,k}$ denote the spectroscopic truth and predicted value, respectively. For redshift ($k = z$), following standard cosmological practice, we adopt the normalized residual $\Delta z_i = (\hat{z}_i - z_i) / (1 + z_i)$ \citep{2008ApJ...686.1503B}, where $z_i$ represents the redshift truth $y_{i,z}$. Unless otherwise specified, all residual-based metrics (RMSE, MAE, $\sigma_{\mathrm{NMAD}}$, and outlier rates) for redshift are computed using $\Delta z_i$, while for all other parameters they are computed using the absolute residual $\Delta y_{i,k}$.

Point-wise statistical performance is quantified using the root mean square error (RMSE) and mean absolute error (MAE):
\begin{align}
\mathrm{RMSE}_k &= \sqrt{\frac{1}{N} \sum_{i=1}^{N} \Delta y_{i,k}^2}, \\
\mathrm{MAE}_k &= \frac{1}{N} \sum_{i=1}^{N} |\Delta y_{i,k}|.
\end{align}

To suppress sensitivity to catastrophic outliers, we employ the normalized median absolute deviation ($\sigma_{\mathrm{NMAD}}$), a robust scatter estimator widely used in observational cosmology \citep{2008ApJ...686.1503B, 2026A&A...714A.121L}:
\begin{equation}
\sigma_{\mathrm{NMAD}, k} = 1.4826 \times \mathrm{median}\left( \left| \Delta y_{i,k} - \mathrm{median}(\Delta y_{i,k}) \right| \right),
\end{equation}
where the normalization factor $1.4826$ ensures equivalence to the standard deviation for a pure Gaussian distribution. While RMSE reflects global prediction errors, $\sigma_{\mathrm{NMAD}}$ specifically quantifies the intrinsic scatter of the core sample population without contamination from extreme tails.

The catastrophic outlier fraction ($\eta_k$, expressed as a percentage) quantifies predictions deviating beyond strict tolerance thresholds:
\begin{equation}
\eta_k = \frac{1}{N} \sum_{i=1}^{N} \mathbb{I}_{i,k} \times 100\%,
\end{equation}
where $\mathbb{I}_{i,k}$ is a binary outlier indicator. For redshift, a prediction is classified as catastrophic if the normalized residual satisfies $|\Delta z_i| > 0.15$, a threshold commonly adopted in photometric redshift studies \citep[e.g.,][]{2008ApJ...686.1503B,2020A&A...644A..31E,2025A&A...702A..74E}. For the remaining physical parameters, we adopt a unified threshold for each parameter, defined as $3\sigma_{\mathrm{NMAD}, k}^{\mathrm{ref}}$. Here $\sigma_{\mathrm{NMAD}, k}^{\mathrm{ref}}$ is a fixed reference scatter determined from the ViTMultiTask with MMD model on the test set. The indicator \(\mathbb{I}_{i,k}\) is formally defined as:
\begin{equation}
\mathbb{I}_{i,k} = \begin{cases}
1, & \text{if } |\Delta z_i| > 0.15 \quad (k = z), \\
1, & \text{if } |\Delta y_{i,k}| > 3\sigma_{\mathrm{NMAD}, k}^{\mathrm{ref}} \quad (k \neq z), \\
0, & \text{otherwise}.
\end{cases}
\end{equation}

In addition to point-wise residual metrics, global distribution fidelity and multi-parameter physical covariances are evaluated using one- and two-dimensional kernel density estimation (1D/2D KDE). For 1D KDE, we compare the predicted and observed spectroscopic-label distributions across all four physical targets to assess overall profile consistency and to examine whether the predicted parameters suffer from distributional distortion, variance shrinkage, or tail truncation. For multi-parameter physical coupling, we construct 2D KDE contour overlays across multiple enclosed probability quantiles, focusing in particular on key astrophysical scaling relations---including the non-linear $D_n 4000$--$\mathrm{H}\delta_A$ evolutionary manifold and the underlying $D_n 4000$--$\log(M_*/\mathrm{M}_\odot)$ distribution---to quantify how accurately the predicted joint probability density reproduces the observed spectroscopic labels.

To quantify the statistical uncertainties of the reported metrics, we estimate $1\sigma$ uncertainties for all continuous metrics (RMSE, MAE, and $\sigma_{\mathrm{NMAD}}$) from 1,000 bootstrap resamples of the test set. For the catastrophic outlier fractions ($\eta$), uncertainties are estimated using Poisson/binomial statistics.

\subsection{Multi-task parameter estimation performance}
\label{sec:performance}

\begin{figure*}[t]
\centering
\includegraphics[width=0.8\textwidth]{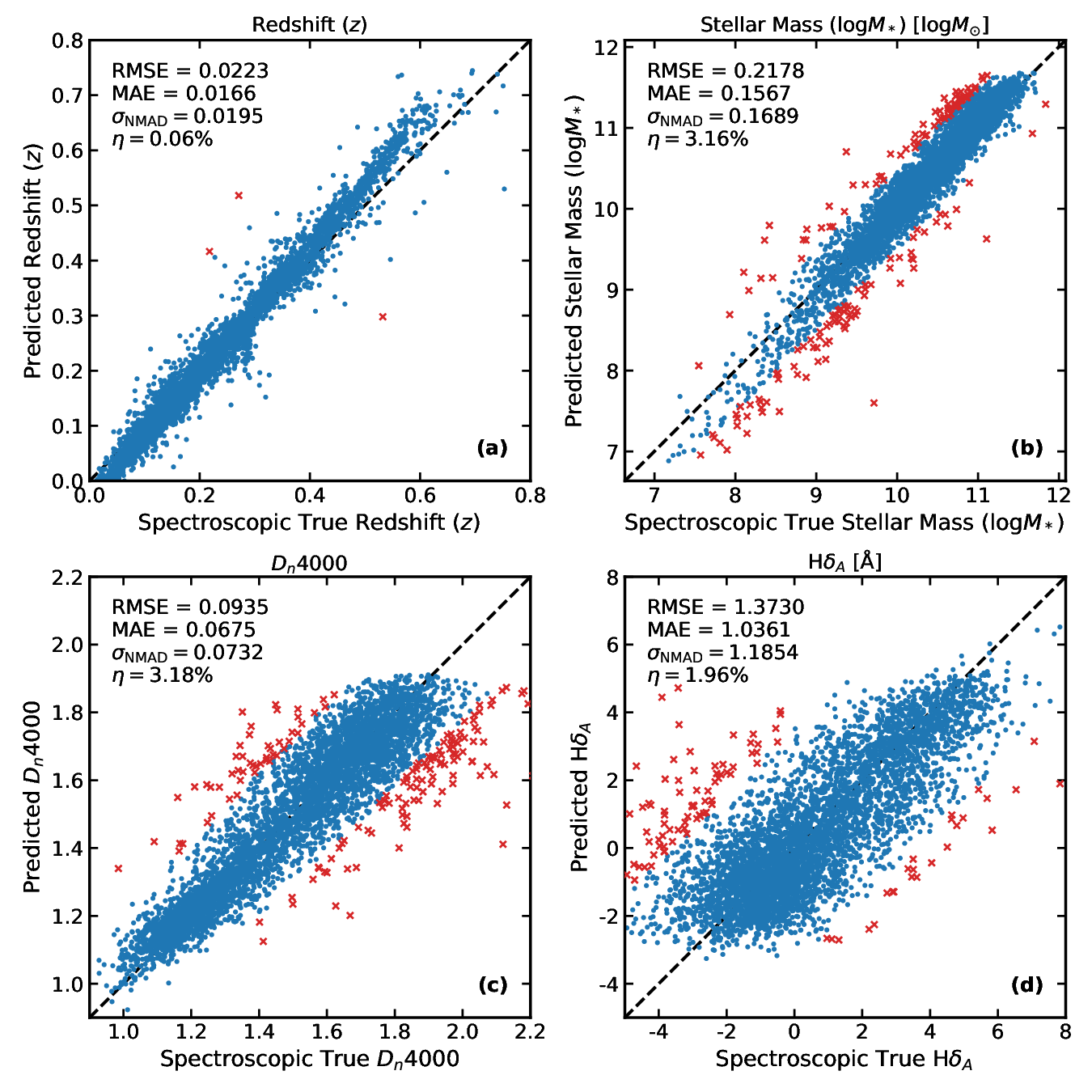}
\caption{Predicted physical parameters versus spectroscopic labels for the ViTMultiTask model on an independent test set of 5,000 galaxies. The four panels display: (a) redshift $z$, (b) stellar mass $\log(M_*/\mathrm{M}_\odot)$, (c) $D_n 4000$ break index, and (d) $\mathrm{H}\delta_A$ equivalent width [\AA]. The black dashed line represents the $1:1$ identity line, blue circles denote well-fitted targets, and red crosses indicate catastrophic outliers. Corresponding evaluation metrics ($\mathrm{RMSE}$, $\mathrm{MAE}$, $\sigma_{\mathrm{NMAD}}$, and outlier fraction $\eta$) are annotated in each panel.}
\label{fig:scatter}
\end{figure*}

Fig.~\ref{fig:scatter} presents the scatter comparisons between predicted physical parameters and spectroscopic labels across the 5,000 test-set galaxies. Quantitative metrics for all parameters are summarized in Table~\ref{tab:ablation}. To establish a consistent evaluation benchmark throughout this study, the unified outlier thresholds for non-redshift physical parameters, $3\sigma_{\mathrm{NMAD}, k}^{\mathrm{ref}}$, are defined using the test-set scatter of the full ViTMultiTask model (with MMD) as the reference, as described in Section~\ref{sec:metrics}.

For redshift $z$, the model achieves the highest point-wise precision among the four tasks, with $\mathrm{RMSE} = 0.0223$, $\mathrm{MAE} = 0.0166$, $\sigma_{\mathrm{NMAD}} = 0.0195$, and a virtually zero catastrophic outlier fraction of $\eta = 0.06\%$. Predictions exhibit no systematic bias across the core range $z \in [0, 0.5]$, with only a marginal overestimation trend at $z \gtrsim 0.5$ where the training sample becomes sparse. This high performance reflects the strong constraining power of five-band photometric colors in breaking color--redshift degeneracies.

Stellar mass is arguably the most dynamically demanding target, spanning nearly five orders of magnitude ($7.0 \lesssim \log(M_*/\mathrm{M}_\odot) \lesssim 12.0$). Despite this broad range, the model delivers $\mathrm{RMSE} = 0.218~\mathrm{dex}$, $\mathrm{MAE} = 0.157~\mathrm{dex}$, $\sigma_{\mathrm{NMAD}} = 0.169~\mathrm{dex}$, and $\eta = 3.16\%$. This compact scatter indicates that ViTMultiTask effectively extracts non-linear mappings linking broadband photometry, spatial morphology, and color gradients to global stellar mass, consistent with well-established scaling relations for low-redshift galaxies \citep{2001ApJ...550..212B,2003ApJS..149..289B,2003MNRAS.341...33K}.

For $D_n 4000$, a classic tracer of mean stellar population age, the model yields $\mathrm{RMSE} = 0.093$, $\mathrm{MAE} = 0.067$, $\sigma_{\mathrm{NMAD}} = 0.073$, and $\eta = 3.18\%$ over the domain $[0.9, 2.2]$. A slight systematic underestimation emerges in the old-galaxy tail ($D_n 4000 > 1.8$), likely reflecting the training set's under-representation of such objects, which biases the network toward the more densely sampled intermediate-age regime. Notably, this deviation is confined to the extreme tail and does not affect the model's overall reliability across the main population ($D_n 4000 \lesssim 1.8$).

The Balmer absorption index $\mathrm{H}\delta_A$ presents the most challenging task, given its extreme sensitivity to recent starbursts, quenching episodes, and potential emission-line infill. Nevertheless, the model attains $\mathrm{RMSE} = 1.373~\mathrm{\AA}$, $\mathrm{MAE} = 1.036~\mathrm{\AA}$, $\sigma_{\mathrm{NMAD}} = 1.185~\mathrm{\AA}$, and $\eta = 1.96\%$ over $[-5.0, 8.0]~\mathrm{\AA}$. This fidelity suggests that the task-specific token mechanism effectively preserves localized spatial information, allowing the network to capture subtle absorption signatures from broadband imaging without the feature-averaging losses inherent to global pooling.

Overall, outliers across all four parameter spaces remain sparse and randomly scattered, with no obvious unphysical systematic trends. These results demonstrate that the proposed architecture robustly handles multi-task regression across physically distinct observables, providing a viable pathway for inferring high-dimensional astrophysical parameters---traditionally requiring high-signal-to-noise spectroscopy---directly from imaging alone.

\subsection{Ablation study: Impact of MMD manifold alignment}
\label{sec:ablation}

\begin{figure*}[t]
\centering
\includegraphics[width=\textwidth]{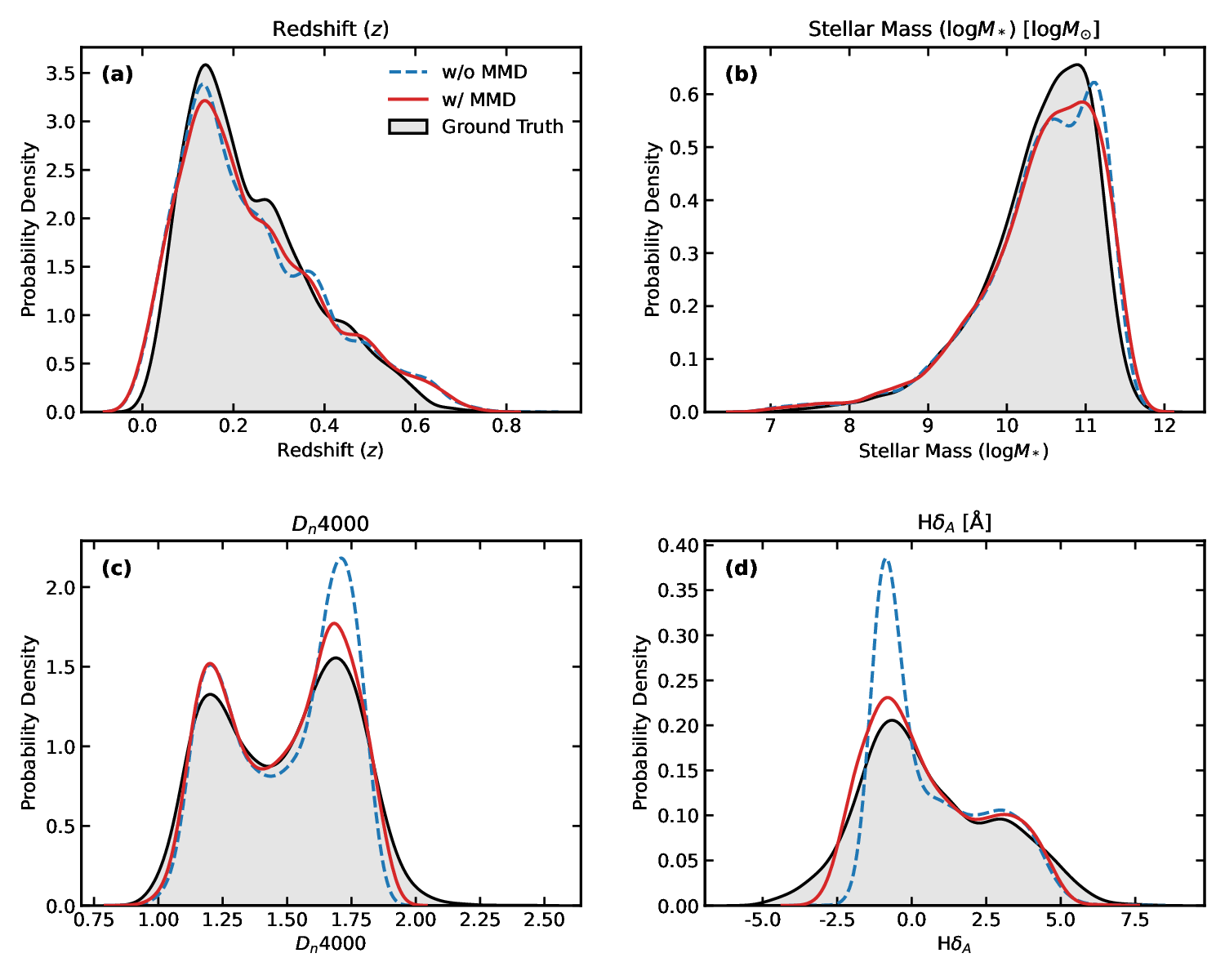}
\caption{One-dimensional kernel density estimates (1D KDE) of predicted physical parameters, comparing predictions without MMD (blue dashed line) and with MMD (red solid line) against the spectroscopic labels (gray filled area). Panels display: (a) redshift $z$, (b) stellar mass $\log(M_*/\mathrm{M}_\odot)$, (c) $D_n 4000$, and (d) $\mathrm{H}\delta_A$ [\AA]. The baseline model without MMD exhibits severe variance shrinkage, resulting in unphysical over-concentrated peaks and truncated distribution tails (notably in panels c and d). In contrast, incorporating MMD better reproduces the variance, multimodal structure, and heavy tails present in the observed spectroscopic-label distributions.}
\label{fig:kde_comp}
\end{figure*}

\begin{figure}[t]
\centering
\includegraphics[width=\columnwidth]{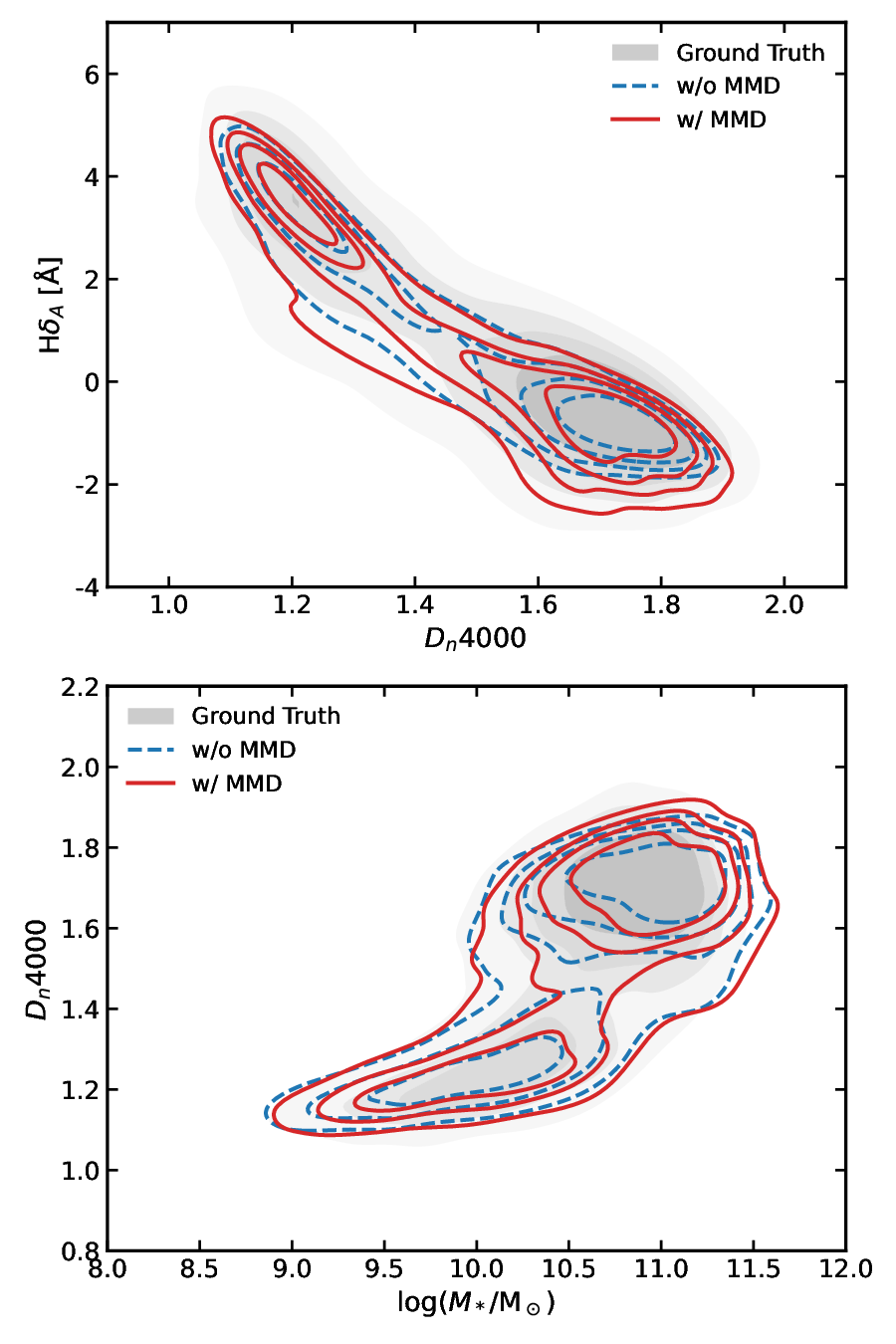}
\caption{Two-dimensional kernel density estimates (2D KDE) evaluating joint physical parameter distribution fidelity. Top panel: $\mathrm{H}\delta_A$ [\AA] versus $D_n 4000$ evolutionary manifold; bottom panel: $D_n 4000$ versus stellar mass $\log(M_*/\mathrm{M}_\odot)$ bimodal distribution. Spectroscopic labels are indicated by gray shaded regions with four discrete intensity gradient levels. Predictions from models trained without MMD (blue dashed contours) and with MMD (red solid contours) are overlaid at the 20\%, 40\%, 60\%, and 80\% enclosed probability quantiles, demonstrating how MMD manifold alignment reproduces the observed parameter couplings in the spectroscopic label and mitigates variance shrinkage.}
\label{fig:kde2d}
\end{figure}

To isolate the role of MMD manifold alignment, we compare two model configurations that are identical in all architectural and training aspects except for the loss function: a baseline trained with pure Huber loss ($\lambda_{\mathrm{MMD}} = 0$), and the full model with MMD regularization ($\lambda_{\mathrm{MMD}} = 1.0$). Throughout this section, agreement with the observed spectroscopic labels (\(P_{\mathrm{obs}}\)) indicates improved fidelity to the empirical target distribution, rather than exact recovery of the underlying intrinsic physical distribution (\(P_{\mathrm{int}}\)). Quantitative results are summarized in Table~\ref{tab:ablation}, and kernel density estimates (KDE) of the predicted distributions against spectroscopic labels are illustrated in Figs.~\ref{fig:kde_comp} and \ref{fig:kde2d}. 

\begin{table*}[htbp]
\centering
\caption{Ablation comparison of model performance with and without MMD loss constraint.}
\label{tab:ablation}
\small
\setlength{\tabcolsep}{6pt}
\renewcommand{\arraystretch}{1.35}
\begin{tabular}{lccccc}
\hline\hline
Parameter & MMD & RMSE & MAE & $\sigma_{\mathrm{NMAD}}$ & $\eta$ (\%) \\
\hline
$z$ & w/o & $0.0225 \pm 0.0007$ & $0.0165 \pm 0.0002$ & $0.0194 \pm 0.0003$ & $0.08 \pm 0.04$ \\
 & w/ & $0.0223 \pm 0.0004$ & $0.0166 \pm 0.0002$ & $0.0195 \pm 0.0004$ & $0.06 \pm 0.03$ \\
\hline
$\log(M_*/\mathrm{M}_\odot)$ (dex) & w/o & $0.211 \pm 0.004$ & $0.152 \pm 0.002$ & $0.163 \pm 0.003$ & $2.94 \pm 0.24$ \\
 & w/ & $0.218 \pm 0.004$ & $0.157 \pm 0.002$ & $0.169 \pm 0.003$ & $3.16 \pm 0.25$ \\
\hline
$D_n 4000$ & w/o & $0.093 \pm 0.002$ & $0.067 \pm 0.001$ & $0.072 \pm 0.002$ & $2.98 \pm 0.24$ \\
 & w/ & $0.094 \pm 0.002$ & $0.067 \pm 0.001$ & $0.073 \pm 0.001$ & $3.18 \pm 0.25$ \\
\hline
$\mathrm{H}\delta_A$ (\AA) & w/o & $1.350 \pm 0.019$ & $1.006 \pm 0.013$ & $1.118 \pm 0.021$ & $2.08 \pm 0.20$ \\
 & w/ & $1.373 \pm 0.018$ & $1.036 \pm 0.012$ & $1.185 \pm 0.022$ & $1.96 \pm 0.20$ \\
\hline\hline
\end{tabular}
\tablefoot{Reported values include continuous metrics alongside their $1\sigma$ statistical uncertainties, estimated via 1,000 bootstrap resamplings for RMSE, MAE, and $\sigma_{\mathrm{NMAD}}$, and Poisson/binomial uncertainties for catastrophic outlier fractions ($\eta$). For a fair comparison, outlier rates ($\eta$) for both models are evaluated using a common threshold ($3\sigma_{\mathrm{NMAD}, k}^{\mathrm{ref}}$) adopted from the ViTMultiTask (w/ MMD) model as the reference.}
\end{table*}

\begin{table*}[htbp]
\centering
\caption{Quantitative evaluation of 1D marginal and 2D joint distribution agreement between predicted parameter distributions and the observed spectroscopic label distribution on the independent test set ($N=5,000$).}
\label{tab:dist_metrics}
\small
\setlength{\tabcolsep}{5pt}
\renewcommand{\arraystretch}{1.25}
\begin{tabular}{lcccccc}
\hline\hline
 & \multicolumn{2}{c}{Wasserstein-1 ($W_1$)} & \multicolumn{2}{c}{Kolmogorov--Smirnov ($D_{\mathrm{KS}}$)} & \multicolumn{2}{c}{1D MMD ($d_{\mathrm{MMD}}^{(1D)}$)} \\
\cline{2-7}
Parameter & w/o MMD & w/ MMD & w/o MMD & w/ MMD & w/o MMD & w/ MMD \\
\hline
$z$ & $0.0141$ & $\mathbf{0.0130}$ & $0.0578$ & $\mathbf{0.0560}$ & $0.1093$ & $\mathbf{0.0973}$ \\
$\log(M_*/\mathrm{M}_\odot)$ (dex) & $\mathbf{0.0610}$ & $0.0633$ & $0.0742$ & $\mathbf{0.0624}$ & $0.1123$ & $\mathbf{0.1023}$ \\
$D_n\,4000$ & $0.0251$ & $\mathbf{0.0124}$ & $0.0666$ & $\mathbf{0.0334}$ & $0.1283$ & $\mathbf{0.0658}$ \\
$\mathrm{H}\delta_A$ (\AA) & $0.3790$ & $\mathbf{0.2123}$ & $0.1468$ & $\mathbf{0.0456}$ & $0.2430$ & $\mathbf{0.0966}$ \\
\hline\hline
\multicolumn{5}{c}{\textbf{2D Joint Distribution MMD Distance ($d_{\mathrm{MMD}}^{(2D)}$)}} & w/o MMD & w/ MMD \\
\hline
\multicolumn{5}{c}{$[\mathrm{H}\delta_A, D_n\,4000]$ Evolutionary Manifold} & $0.2321$ & $\mathbf{0.1542}$ \\
\multicolumn{5}{c}{$[D_n\,4000, \log(M_*/\mathrm{M}_\odot)]$ Bimodal Sequence} & $0.1206$ & $\mathbf{0.0941}$ \\
\hline\hline
\end{tabular}
\tablefoot{Lower values indicate better agreement with the observed spectroscopic label distribution. Bold entries highlight the better-performing model configuration for each metric.}
\end{table*}

Unconstrained point-wise losses (e.g., Huber loss) encourage network predictions toward conditional expectations $\mathbb{E}[Y \mid X]$, inducing "variance shrinkage" and regression to the mean. As summarized in Table~\ref{tab:ablation}, the baseline model without MMD achieves marginally lower point-wise residual metrics for stellar mass and spectral indices (e.g., $\mathrm{RMSE} = 0.211~\mathrm{dex}$ for $\log(M_*/\mathrm{M}_\odot)$ and $0.093$ for $D_n4000$). However, as demonstrated in Fig.~\ref{fig:kde_comp}, this marginal numeric gain comes at the cost of severe distribution distortion. Without MMD, the predicted distributions artificially narrow, producing unphysical sharp peaks around modal nodes (e.g., $D_n4000 \approx 1.70$, $\mathrm{H}\delta_A \approx -0.80~\mathrm{\AA}$) and truncating distribution tails in extreme parameter regimes (e.g., failing to capture the extended negative tail of $\mathrm{H}\delta_A < -2.0~\mathrm{\AA}$ and over-concentrating stellar mass near $\log(M_*/\mathrm{M}_\odot) \approx 11.1$).

By contrast, the multi-scale MMD loss encourages the predicted joint distribution $P(\hat{\mathbf{y}})$ to align with the target joint distribution $Q(\mathbf{y})$ within the RKHS framework. This effectively mitigates variance shrinkage, better preserving the multimodal shapes, predictive dispersion, and heavy-tail structures observed in the spectroscopic labels, which are critical for astrophysical interpretation. Under this standardized benchmark, the MMD-regularized model introduces no significant degradation in point-estimate accuracy (RMSE varies by $\le 3.5\%$ across all physical quantities and remains statistically consistent with the baseline) and yields comparable outlier rates. Specifically, for the stellar-population tracer $\mathrm{H}\delta_A$, the MMD model achieves a slightly lower outlier fraction than the baseline ($1.96\%$ vs. $2.08\%$), although the difference is not statistically significant. For global parameters such as stellar mass and $D_n4000$, the outlier fractions remain consistent within Poisson/binomial uncertainties ($3.16\%$ vs. $2.94\%$ for $\log(M_*/\mathrm{M}_\odot)$; $3.18\%$ vs. $2.98\%$ for $D_n4000$; MMD vs. baseline). Overall, MMD delivers clear gains in global distribution fidelity---a property of particular value for cosmological applications such as galaxy stellar mass functions and galaxy evolution modeling.

To further evaluate the model's fidelity in preserving multi-dimensional stellar population correlations, Fig.~\ref{fig:kde2d} presents the predicted 2D joint probability density contours for the $\mathrm{H}\delta_A$ versus $D_n 4000$ evolutionary manifold (top panel) and the $D_n 4000$ versus $\log(M_*/\mathrm{M}_\odot)$ sequence (bottom panel). The spectroscopic labels (gray shaded regions) clearly delineates the characteristic bimodal distribution of galaxy populations, spanning from young, star-forming galaxies (low $D_n 4000$, high $\mathrm{H}\delta_A$, lower mass) in the upper-left region of the joint parameter spaces to old, quenched systems (high $D_n 4000$, negative $\mathrm{H}\delta_A$, higher mass) in the quiescent sequence.

While both model configurations capture the primary bimodal structure, the benefits of MMD regularization become clear when examining the two joint spaces separately. For the $\mathrm{H}\delta_A$--$D_n4000$ evolutionary manifold (top panel), the spectroscopic labels traces a characteristic curved sequence from post-starburst galaxies (high $\mathrm{H}\delta_A$, low $D_n4000$) down to old, quiescent systems (negative $\mathrm{H}\delta_A$, high $D_n4000$). The baseline model trained without MMD (blue dashed contours) captures the central ridge but compresses the outer low-density envelopes, struggling to reproduce the extended tails at both ends of the evolutionary track. Incorporating MMD regularization (red solid contours) expands the predicted contours outward across the $20\% \text{--} 80\%$ enclosed probability quantiles, achieving substantially better alignment with the observed distribution across the full dynamic range---particularly in recovering the high-$\mathrm{H}\delta_A$ tail that traces recent starburst activity.

For the $D_n 4000$--$\log(M_*/\mathrm{M}_\odot)$ joint distribution (bottom panel), the observed labels exhibit a well-known bimodal structure: a low-mass star-forming cloud ($\log(M_*/\mathrm{M}_\odot) \lesssim 10.0$, low $D_n 4000$) and a high-mass red sequence ($\log(M_*/\mathrm{M}_\odot) \gtrsim 10.5$, high $D_n 4000$), connected by a relatively sparsely populated green valley. The baseline model again displays variance shrinkage, over-concentrating predictions around the modal peaks and under-representing the low-density bridge and the low-mass end of the star-forming sequence. Incorporating MMD expands the predicted contours toward the observed distribution boundaries, recovering the full bimodal geometry and providing more faithful coverage of the transitional green-valley population. Taken together across both joint spaces, the MMD-constrained model can effectively capture the non-linear evolutionary manifold between Balmer absorption, $4000~\mathrm{\AA}$ break strength, and stellar mass, while preserving the coupling between these parameters observed in the spectroscopic labels that point-wise losses alone tend to smooth over.

Regarding density estimation, all 1D and 2D KDE distributions in Figs.~\ref{fig:kde_comp} and \ref{fig:kde2d} are generated using Gaussian kernels. For 1D KDEs, bandwidths are selected using Scott's rule of thumb. For 2D KDEs, a minor adjustment ($\texttt{bw\_adjust} = 0.8$) is applied relative to Scott's rule to resolve fine bivariate manifold structures. Importantly, the same smoothing parameters and density thresholds are applied consistently to both the spectroscopic labels and model predictions, ensuring that the visual comparison is not dominated by bandwidth selection.

To quantitatively supplement the kernel density visualizations, we evaluate distribution alignment for all individual parameters and selected joint spaces using three metrics: the Wasserstein-1 distance ($W_1$; \citealt{arjovsky2017wasserstein,2025ApJS..277...22L}), the Kolmogorov--Smirnov statistic ($D_{\mathrm{KS}}$), and the multivariate MMD distance ($d_{\mathrm{MMD}}$) in RKHS (Table~\ref{tab:dist_metrics}). We denote MMD values for 1D marginals and 2D joint distributions as $d_{\mathrm{MMD}}^{(1D)}$ and $d_{\mathrm{MMD}}^{(2D)}$, respectively. Compared with the baseline model without MMD, incorporating MMD regularization yields substantial reductions in distribution discrepancies, particularly for spectroscopically sensitive features. For $D_n4000$, $W_1$ decreases by 50.6\% ($0.0251 \to 0.0124$) and $d_{\mathrm{MMD}}^{(1D)}$ decreases by 48.7\%. For $\mathrm{H}\delta_A$, $W_1$ decreases by 44.0\% ($0.3790 \to 0.2123$), $D_{\mathrm{KS}}$ decreases by 68.9\% ($0.1468 \to 0.0456$), and $d_{\mathrm{MMD}}^{(1D)}$ decreases by 60.2\%. For the overall shape of $z$ and stellar mass, $D_{\mathrm{KS}}$ and $d_{\mathrm{MMD}}^{(1D)}$ also show consistent improvements, while their $W_1$ distances remain essentially unchanged (e.g., \(0.0141 \to 0.0130\) for \(z\) and \(0.0610 \to 0.0633\) for \(\log(M_*/\mathrm{M}_\odot)\)). In the 2D joint spaces, MMD regularization reduces $d_{\mathrm{MMD}}^{(2D)}$ by 33.6\% for the $[\mathrm{H}\delta_A, D_n4000]$ evolutionary manifold and by 22.0\% for the $[D_n4000, \log(M_*/\mathrm{M}_\odot)]$ bimodal sequence, indicating improved consistency with the observed distribution contours.

A slight residual gap persists between the outermost MMD contours and the observed distribution boundary. This minor offset is primarily attributed to two factors. First, intrinsic observational noise and measurement uncertainties in the spectroscopic estimates introduce additional broadening into the empirical label distribution, making its outermost boundaries inherently difficult to reproduce exactly from imaging alone. Second, the joint optimization of point-wise Huber regression and global MMD regularization entails a fundamental trade-off: prioritizing distribution alignment can slightly compromise per-sample point accuracy, and vice versa. One promising direction for future work would be to directly incorporate measurement uncertainties into the distribution alignment loss, which may help further narrow this residual gap.

%%%%%%%%%%%%%%%%%%%%%%%%%%%%%%%%%%%%%%%%%%%%%%%%%%%%%%%%%%%%%%

%%%%%%%%%%%%%%%%%%%%%%%%%%%%%%%%%%%%%%%%%%%%%%%%%%%%%%%%%%%%%%
\section{Summary and Conclusions}
\label{sec:conclusions}

We have presented ViTMultiTask, a multi-task deep learning framework that integrates a vision transformer (ViT) backbone with task-specific tokens for feature decoupling, optimized via a multi-scale maximum mean discrepancy (MMD) loss for joint distribution alignment. The framework is designed to directly infer photometric redshifts ($z$) and key stellar population parameters ($\log(M_*/\mathrm{M}_\odot)$, $D_n 4000$, and $\mathrm{H}\delta_A$) from Subaru HSC five-band ($g,r,i,z,y$) imaging cutouts. By encouraging alignment between predicted and observed spectroscopic-label distributions in a reproducing kernel Hilbert space (RKHS), the MMD loss effectively mitigates the variance shrinkage and distribution distortion inherent to conventional point-wise regression losses.

Evaluated on 83,846 galaxies cross-matched between HSC PDR2 and DESI Data Release 1 BGS, the network demonstrates robust cross-modal mapping from broadband imaging to high-resolution spectroscopic diagnostics. On an independent test set of 5,000 galaxies, the pipeline achieves $\sigma_{\mathrm{NMAD}} = 0.020$ ($\mathrm{RMSE} = 0.022$, $\eta = 0.06\%$) for redshift and $\sigma_{\mathrm{NMAD}} = 0.169~\mathrm{dex}$ ($\mathrm{RMSE} = 0.218~\mathrm{dex}$) for stellar mass. For the spectroscopic indices, $D_n 4000$ yields $\mathrm{RMSE} = 0.094$, $\sigma_{\mathrm{NMAD}} = 0.073$ ($\eta = 3.18\%$), while $\mathrm{H}\delta_A$ achieves $\mathrm{RMSE} = 1.373~\mathrm{\AA}$, $\sigma_{\mathrm{NMAD}} = 1.185~\mathrm{\AA}$ ($\eta = 1.96\%$). These results indicate that the ViT backbone effectively extracts spatially resolved morphological and color gradient information directly from multi-band imaging.

Ablation experiments highlight the central role of MMD-based distribution alignment. Although the baseline model trained with Huber loss alone achieves marginally lower point-wise residuals for certain parameters, it exhibits pronounced variance shrinkage and fails to reproduce the observed bimodal structures and heavy tails in the empirical distributions of the spectroscopic labels. In 1D kernel density estimates, the baseline artificially narrows the predicted distributions, producing over-concentrated peaks around modal nodes and truncating extreme tails (e.g., the extended negative tail of $\mathrm{H}\delta_A$ and the high-mass end of $\log M_*$). In 2D joint spaces, it compresses the $\mathrm{H}\delta_A$--$D_n 4000$ evolutionary manifold and under-represents the low-mass blue cloud and green-valley transition in the $D_n 4000$--$\log(M_*/\mathrm{M}_\odot)$ plane. Incorporating MMD regularization ($\lambda_{\mathrm{MMD}} = 1.0$) incurs no significant degradation in point-wise precision (changes in RMSE of $\le 3.5\%$, statistically consisitent with the baseline) while clearly improving global distribution fidelity. For large-scale cosmological surveys, improving agreement with observed global parameter distributions (while accounting for target sample uncertainties) is more crucial for reducing bias in statistical inference than solely minimizing per-sample residuals.

Minor systematic underestimations persist at high $D_n 4000$ ($> 1.8$) and at the highest redshifts ($z \gtrsim 0.5$), reflecting the intrinsic sparsity of massive, quenched systems in the training sample. Future work will explore self-supervised pre-training strategies (such as Masked Autoencoders, MAE) to enhance feature extraction for rare populations and further improve robustness in data-sparse regions. More broadly, applying this manifold alignment paradigm to upcoming wide-field surveys---such as CSST, Euclid, and LSST---offers a promising pathway toward reliable, large-scale galaxy property estimation for the billions of photometrically observed targets that will lack spectroscopic coverage, fully addressing the spectroscopically limited nature of the deep imaging era.

%%%%%%%%%%%%%%%%%%%%%%%%%%%%%%%%%%%%%%%%%%%%%%%%%%%%%%%%%%%%%%
\begin{acknowledgements}
Z.L. acknowledges the support from the National Natural Science Foundation of China (Grant No. 12573009) and the scientific research grants from the China Manned Space Project with Grant Nos. CMS-CSST-2025-A07 and CMS-CSST-2025-A05. This work is also supported by the National Natural Science Foundation of China under Grant No. 12141302.
\end{acknowledgements}

\section*{Data availability}

The code for ViTMultiTask supporting the findings of this study is publicly available on GitHub at \url{https://github.com/zjluo-code/ViTmultitask}. The HSC imaging data and spectroscopic redshift labels were derived from \citet{2025arXiv251022527H} (\url{https://zenodo.org/records/16989593}). The target physical parameters, including stellar mass ($\log(M_*/\mathrm{M}_\odot)$), $D_n 4000$, and $\mathrm{H}\delta_A$, were obtained from ScienceDB \citep{2026arXiv260719162L} (\url{http://www.doi.org/10.57760/sciencedb.35360}). The curated training, validation, and test datasets compiled for this work are archived on Zenodo at \url{https://doi.org/10.5281/zenodo.24154596} and will be made publicly accessible upon publication.

\bibliographystyle{aa}  
\bibliography{ref} 

% Please note that we have included the references below in
% order to compile the document, but we ask you to:

\end{document}